\documentclass[epj,final]{svjour}
\usepackage{graphicx}
\DeclareGraphicsExtensions{.eps}
\usepackage{amsmath}
\usepackage[latin1]{inputenc}
\newlength{\figwidth}
\begin{document}
\title{Residual conductance of correlated one-dimensional nanosystems:
A numerical approach}
\titlerunning{Conductance of correlated nanosystems}

\author{%
Rafael A.\ Molina\inst{1}
\and 
Peter Schmitteckert\inst{2}
\and
Dietmar Weinmann\inst{3}
\and
Rodolfo A.\ Jalabert\inst{3}
\and
Gert-Ludwig Ingold\inst{4}
\and
Jean-Louis Pichard\inst{1,5}
}
\authorrunning{R.\ A.\ Molina \textit{et al}.}
\institute{%
CEA/DSM, Service de Physique de l'Etat Condens{\'e},
Centre d'Etudes de Saclay, 91191 Gif-sur-Yvette, France 
\and
Institut f{\"u}r Theorie der Kondensierten Materie, 
Universit{\"a}t Karlsruhe, 76128 Karlsruhe, Germany
\and
Institut de Physique et Chimie des Mat{\'e}riaux de Strasbourg,
UMR 7504 (CNRS-ULP), 23 rue du Loess, BP 43, 67034 Strasbourg Cedex 2, 
France 
\and 
Institut f{\"u}r Physik, Universit{\"a}t Augsburg,
Universit{\"a}tsstra{\ss}e 1, 86135 Augsburg, Germany 
\and Laboratoire de Physique Th\'eorique et Mod{\'e}lisation,
Universit{\'e} de Cergy-Pontoise, 95031 Cergy-Pontoise Cedex, France}

\date{\today}

\abstract{
We study a method to determine the residual conductance of a correlated
system by means of the ground-state properties of a large ring composed
of the system itself and a long non-interacting lead. The transmission 
probability through the interacting region and thus its residual conductance 
is deduced from the persistent current induced by a flux threading the ring. 
Density Matrix Renormalization Group techniques are employed to obtain 
numerical results for one-dimensional systems of interacting spinless 
fermions. As the flux dependence of the persistent current for such a 
system demonstrates, the interacting system coupled to an infinite 
non-interacting lead behaves as a non-interacting scatterer, but with 
an interaction dependent elastic transmission coefficient. The scaling 
to large lead sizes is discussed in detail as it constitutes a crucial 
step in determining the conductance. Furthermore, the method, which so 
far had been used at half filling, is extended to arbitrary filling 
and also applied to disordered interacting systems, where it is found 
that repulsive interaction can favor transport.
}
\maketitle

\section{Introduction}

Large experimental activities have recently been devoted to the study of the
conductance of low-dimensional nanosystems like molecules, atomic chains, 
nanotubes, and quantum wires 
\cite{Yao,smit03,Nygard,Joaquim,molecular_electronics} with sizes typically of 
the order of the electronic Fermi wavelength. Since the screening of the 
Coulomb interaction in such systems is less effective than in three 
dimensions, electronic correlations can no longer be neglected with respect to 
kinetic effects. In some of the systems mentioned, the Luttinger
liquid behavior \cite{maslov,safi} is relevant and might 
influence the transport properties.

The correlations become particularly relevant for low temperature electronic 
transport properties like the residual conductance and the interpretation
of the experimental data requires a good understanding of transport through
a region with strong correlations. However, this turns out to be a demanding
task and various attempts have been made in this direction
\cite{meirw92,datta}.

The purpose of the present work is to contribute to the fundamental
problem of transport through correlated nanostructures by studying a novel
approach where the conductance is obtained from thermodynamic properties of 
a ring consisting of the nanosystem and a long lead. Such an embedding method 
has been actively pursued in the last few years
\cite{favan98,sushk01,molin03,meden03,meden03a,rejec03,rejec03a}.
Here, we critically study its hypotheses and consequences in order to 
put it on a firm theoretical basis. 

A powerful concept which was used for studying coherent transport through 
non-interacting systems is the Landauer-B\"uttiker formalism 
\cite{Landauer,Buttiker} which formulates a scattering problem between 
electron reservoirs. Although the electrons in the reservoirs interact, 
their density is very high such that the Coulomb energy to kinetic energy 
is small and they can be replaced by non-interacting quasiparticles. 
Hence, the reservoirs are well described by a Fermi distribution 
characterized by a temperature and a chemical potential. Within the 
scattering approach, the dimensionless residual conductance 
$g$ (in units of $e^2/h$) is given by the elastic transmission 
probability $\vert t(E_\mathrm{F})\vert^2$ at the Fermi energy 
$E_\mathrm{F}$.

The situation becomes more complicated if electron-electron interaction is
present in the scattering region because the passage of electrons may lead to
the creation of excitations. However, for temperatures smaller than the 
characteristic excitation energy of the nanosystem, the idea of the 
Landauer-B\"uttiker formalism still applies \cite{meirw92} because
all accessible states in the reservoir with an energy lower than
the excitation energy are occupied. Inelastic processes are then forbidden.
On the other hand, it remains non-trivial to determine
the elastic transmission probability through a correlated system. Green
function methods, while being conceptually adequate, require knowledge of
the excited states and may become numerically quite involved.

\begin{figure}[tb]
\centerline{\includegraphics[width=0.7\figwidth]{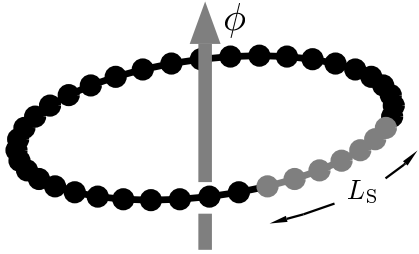}}
\vspace{2mm}
\caption[]{\label{fig:system} The system considered within the embedding
approach is a one-dimensional ring consisting of an interacting region (grey) 
of length $L_\mathrm{S}$ and a non-interacting lead (black) of length
$L_\mathrm{L}$. The ring is threaded by an Aharonov-Bohm flux $\phi$.}  
\end{figure}

An alternative approach consists in considering the ground state
properties of a ring formed by the system of interest, which we will refer
to as correlated system or nanosystem, and a very long non-interacting lead 
as depicted in Fig.~\ref{fig:system}. Within this embedding method, the 
relevant information about the conductance can be extracted by means of a 
flux threading the ring, which gives rise to a flux dependence of the 
ground-state energy and thus to a persistent current. This setup accounts 
for two important physical ingredients of coherent transport. First, the 
flux dependence of the ground-state energy provides information about 
extended states in the interacting region. Second, the two contacts between 
system and lead allow to transfer electrons into the system. This is an 
essential point in the description of conductance \cite{berkovits}, 
which is not present when the persistent current is calculated 
for a correlated system without auxiliary lead.

Favand and Mila used the above described approach to compare, within a model
of spinless fermions, the tunneling conductance of molecules with a 
Mott-Hubbard gap and of molecules with a dimerization gap \cite{favan98}.
Sushkov used the same idea for a study of the $0.7 e^2/h$ anomaly observed in 
quantum point contacts \cite{sushk01,sushk03}. However, an important 
difference with respect to Ref.~\cite{favan98} is that he kept the 
interaction in the leads within the Hartree-Fock approximation. As the
present authors have emphasized \cite{molin03}, the extrapolation to infinite 
lead length can only yield meaningful results if no interaction is present 
in the auxiliary lead. Other important aspects discussed in 
Ref.~\cite{molin03} are the relevance of the contacts, the 
oscillation of the conductance as a function of the number of sites in the 
interacting region, and the role of static disorder. Meden and Schollw{\"o}ck 
compared the results obtained within this approach to those of a perturbative 
functional 
renormalization group and showed that both give the same results at small 
values of the interaction strength, verifying scaling laws associated 
with Luttinger liquid behavior \cite{meden03,meden03a}. 
Rejec and Ram{\v s}ak tested the method, comparing its prediction with 
previous results for transport through single and double quantum dots. 
They presented a generalization to systems without time-reversal 
symmetry, using as an example a nanosystem which itself forms an 
Aharonov-Bohm ring \cite{rejec03,rejec03a}. 

An approach related to the embedding method has recently been 
proposed by Chiappe and Verg{\'e}s \cite{chiappe} in which the
nanosystem and a small part of the leads are diagonalized exactly. 
In a second step, this subsystem is attached to semi-infinite leads 
and Green functions are employed to numerically calculate the conductance.
The conductance through a one-dimensional interacting
spin-system coupled to non-interacting leads was also studied by Louis and 
Gros by means of a Monte-Carlo based method \cite{louis}. 

The relationship between the conductance and the persistent current of a 
large ring has only been proven for non-interacting scattering systems.
No rigorous proof has so far been put forward once electronic 
correlations are present in the scattering region. However, the 
conductance obtained by means of the embedding method satisfies all 
basic requirements and reproduces the correct behavior in various 
limiting cases. Moreover, in this work we demonstrate numerically for the 
one-dimensional case that in the limit of very large
ring size, the effect of an interacting scatterer on the persistent current 
can be described by the amplitude of a transmission probability 
characterizing a $2 \times 2$ transfer matrix. Thus, transport through 
an interacting region can be understood as a non-interacting scatterer 
with interaction dependent parameters.

The remainder of the paper is organized as follows. 
In Section~\ref{sec:flux-dependence}, we use Density Matrix Renormalization 
Group (DMRG) techniques to calculate the flux dependence of the persistent 
current through a ring composed of an interacting region
and a non-interacting auxiliary lead in the limit where the latter becomes 
very long. It is found that this flux dependence reproduces the one expected
for a non-interacting ring of equal length interrupted by a scatterer which
can be characterized by a transfer matrix.

In the absence of Luttinger-like correlations in the ring it is meaningful to 
consider the limit of a very long auxiliary lead. In 
Section~\ref{sec:condfromtrans} we will explain how the extrapolation to 
infinite circumference can be performed in order to extract the 
interaction-dependent transmission coefficient and thus the conductance. For 
this scaling analysis, we make use of the charge stiffness instead of the 
persistent current, because it provides us with the same information but 
requires less numerical effort. Specific attention will be paid to the case 
of resonances, which appear when the coupling between system and leads is 
small and where the extrapolation has to be done with particular care.

In the literature, the embedding method has so far been discussed only for 
the case of half filling. In Section~\ref{sec:arbitrary-filling} we will 
present an extension to arbitrary filling. The important point is to choose 
the appropriate compensating background potential which ensures the correct
charge density in the system even in the presence of interactions. While
at half filling, it is straightforward to define the compensating potential
from particle-hole symmetry, a self-consistent procedure is required away
from half filling. In Section~\ref{sec:disorder}, we employ this new method
to demonstrate that strong repulsive interactions can favor zero-temperature
transport through strongly disordered systems.

We present our conclusions and perspectives in Section~\ref{sec:summary}. In
Appendix \ref{sec:noninteracting} we address the flux dependence of the 
ground state for a ring containing a local non-interacting scatterer,
and obtain the asymptotic values and the finite size corrections to the 
charge stiffness. In Appendix \ref{sec:super} we extend the approach to 
superconducting nanosystems and verify that the known behavior resulting 
from Andreev scattering at the two extremities of a superconducting 
nanosystem is reproduced. This illustrates the validity of the studied 
embedding method in an extreme limit where an attractive electron-electron 
interaction has dramatic effects.

\section{Flux dependence of the persistent current for large rings
  with a small scattering region}
\label{sec:flux-dependence}

The aim of this section is to demonstrate that the transport properties of
an interacting region can be described as a non-interacting scattering 
problem with interaction dependent parameters. We start by considering 
the setup shown in Fig.~\ref{fig:system} which will be employed to study 
the transport properties of a one-dimensional system of length 
$L_\mathrm{S}$. This system may contain a scattering potential and, 
possibly, electron-electron interaction may be present there.

The system is contacted by the two ends of an auxiliary
one-dimensional lead of length $L_\mathrm{L}$ so that a ring of total 
length $L=L_\mathrm{S}+L_\mathrm{L}$ is formed. From this setup, 
transmission properties of the system can only be deduced if 
Luttinger liquid correlations \cite{kane_fisher} in the one-dimensional 
ring are absent. Therefore it is crucial that in the auxiliary lead no 
electron-electron interaction may be present. Not only, this allows to 
avoid Luttinger liquid correlations, but the electrons of the combined 
ring form a Fermi liquid in the limit of infinite lead length. According 
to Sushkov, one can give a general argument for 1d spinless fermions 
on a ring demonstrating that they form a Fermi liquid though interactions 
act in a region of the ring \cite{sushkovprivate}, as far as it remains 
finite while the non-interacting lead becomes infinite. This is corroborated 
by our numerical findings presented below.  

Information about the transmission amplitude $|t(E_\mathrm{F})|$ at the Fermi 
energy $E_\mathrm{F}$ can be obtained by means of a magnetic flux $\phi$ 
threading the ring. For convenience, we introduce the dimensionless flux 
$\Phi=2\pi\phi/\phi_0$ where $\phi_0=h/e$ is the flux quantum. The many-body 
ground state energy $E_0$ of the ring will oscillate with period $\phi_0$ as 
a function of the flux. The magnetic flux threading the ring breaks the 
symmetry between left and right moving electrons and thus gives rise to a 
persistent current $J(\Phi)$, which at zero temperature is given by 
$J(\Phi)=-\partial E_0/ \partial \phi$. 

For non-interacting scatterers, the persistent current $J(\Phi)$ decreases 
like $1/L$ for large circumference $L$ of the ring. The leading contribution 
is found to read \cite{gogolin}
\begin{equation}
J(\Phi) = -\frac{ev_\mathrm{F}}{\pi L} \frac{\mathrm{Arccos}
              \big(|t(k_\mathrm{F})| \cos(\Phi)\big)}
     {\sqrt{1- |t(k_\mathrm{F})|^2\cos^2(\Phi)}} |t(k_\mathrm{F})| \sin(\Phi) 
\label{eq:jfluxodd}
\end{equation}
for an odd number of particles and
\begin{equation}
J(\Phi) = \frac{ev_\mathrm{F}}{\pi L} \frac{\mathrm{Arccos}
              \big(|t(k_\mathrm{F})| \cos(\Phi -\pi)\big)}
      {\sqrt{1- |t(k_\mathrm{F})|^2 \cos^2(\Phi)}} |t(k_\mathrm{F})| \sin(\Phi) 
\label{eq:jfluxeven}
\end{equation}
for the case of an even number of particles in the ring. By Arccos, we denote 
the principal branch of the inverse cosine function which takes values in the 
interval $[0,\pi]$.  The derivation of these results is outlined in 
Appendix \ref{sec:noninteracting}.

The persistent currents (\ref{eq:jfluxodd}) and (\ref{eq:jfluxeven})
depend on the properties of the non-interacting scatterer only through its
transmission probability $\left|t(E_{\rm F})\right|^2$ at the Fermi 
energy. This important feature allows us to determine the transmission 
probability and thus the residual conductance of the system from the 
persistent current of the composed ring. 
The relation becomes particularly simple for $\Phi=\pi/2$, where 
the transmission coefficient at the 
Fermi energy can be expressed as \cite{Gefen,favan98,sushk01}
\begin{equation}\label{eq:transpc}
|t(E_\mathrm{F})|^2=\left(\frac{J(\pi/2)}{J^0(\pi/2)}\right)^2 \, .
\end{equation}
Here, $J^0$ is the persistent current for a clean ring of length
$L$. 

We now turn to an interacting nanosystem and demonstrate numerically that, in
the limit of an infinitely long lead, the flux dependence of the persistent 
current is of the same form as in the non-interacting case of 
Eqs.\ (\ref{eq:jfluxodd}) and (\ref{eq:jfluxeven}). 
The interaction thus enters the result only through the transmission 
coefficient $\left|t(E_{\rm F},U)\right|^2$. 

Specifically, we have performed direct numerical calculations of the 
persistent current for a tight-binding model with $N$ interacting spinless 
fermions on $L$ sites described by the Hamiltonian  
\begin{equation}\label{eq:hamil}
H = -t\sum_{i=1}^{L}(c^\dagger_i c^{\phantom{\dagger}}_{i-1} 
+ c^\dagger_{i-1}c^{\phantom{\dagger}}_i)
+ \sum_{i=2}^{L_\mathrm{S}}U\left[n_i-V_+\right]
\left[n_{i-1}-V_+\right] \, .
\end{equation}
The hopping amplitude $t$ between nearest neighbors will be
set to 1 and thus defines our energy scale.  
$c^{\phantom{\dagger}}_i$ ($c^\dagger_i$) is the annihilation (creation)
operator at site $i$, $n_i = c^\dagger_i c^{\phantom{\dagger}}_i$ is
the number operator, and the flux enters through the boundary condition 
$c_0 = \exp(\mathrm{i}\Phi )c_L$. The length scale is given by the lattice
spacing and the interaction acts between nearest neighbors inside 
the sample (sites $i=1$ to $L_\mathrm{S}$), but vanishes in the lead. To 
avoid depletion of electrons in the sample due to the repulsive interaction, 
we introduce a compensating potential $V_+$ that acts as a positive 
background charge and ensures the local charge neutrality. For a half-filled
ring, the compensating potential is equal to the filling factor
$\nu=N/L$. Thus,
\begin{equation}\label{eq:ham}
V_+\left(\nu=\frac{1}{2}\right)=\frac{1}{2}
\end{equation}
guarantees particle-hole symmetry even in the presence of
interactions. Outside half filling, this symmetry is broken and the 
compensating potential $V_+$ becomes a function of $U$, $N$, 
$L_\mathrm{S}$ and $L_\mathrm{L}$ as we will discuss in 
Section~\ref{sec:arbitrary-filling}.

For the model (\ref{eq:hamil}) with system size $L_\mathrm{S}=6$ 
and at half filling, we have numerically determined the persistent 
current $J(\Phi)$ supported by the 
ground state for various values of $L_\mathrm{L}$ by means of a complex DMRG 
algorithm. With this implementation, we are able to treat not only the flux 
values $\Phi=0$ and $\Phi=\pi$ used in \cite{molin03}, where the Hamiltonian 
(\ref{eq:hamil}) can be represented by a real matrix, but also the general 
case of arbitrary flux where the matrix becomes complex. In order to 
determine the persistent current, we directly evaluate the current operator 
for the ground state, thereby avoiding the potentially difficult procedure 
of taking numerically the derivative of $E_0(\Phi)$.

\begin{figure}
\centerline{\includegraphics[width=\figwidth]{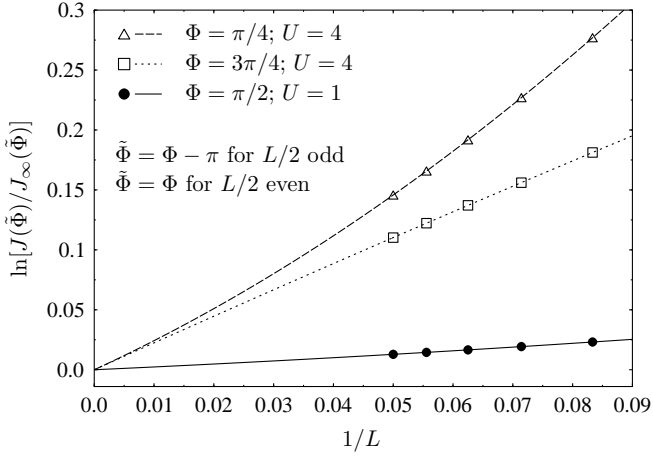}}
\vspace{2mm}
\caption[]{\label{fig:pc-scaling} 
The scaling of the persistent current with the total length $L$ is performed
for a ring with system size $L_\mathrm{S}=6$ at half filling for several
values of the flux $\Phi$ and the interaction strength $U$. The persistent
current $J(\Phi)$ is depicted for even particle numbers $N=6,8,10$ while for
odd particle numbers $N=7, 9$, results for $J(\tilde{\Phi})$ 
with $\tilde{\Phi}=\Phi-\pi$ are shown. The extrapolation $L\to\infty$ has been
performed by means of fits to second-order polynomials in $1/L$.}
\end{figure}

The length dependence of the persistent current and the extrapolation to 
infinite lead length is shown in Fig.~\ref{fig:pc-scaling} for particle
numbers $N=L/2$ between 6 and 10. Motivated by the 
symmetry
\begin{equation}\label{eq:pc-symmetry}
J(\Phi;N\ \mathrm{odd})=J(\Phi-\pi;N\ \mathrm{even})\, ,
\end{equation}
valid in the non-interacting case according to (\ref{eq:jfluxodd}) and
(\ref{eq:jfluxeven}), we plot the interacting results corresponding to even 
and odd $N$ at flux values $\Phi$ and $\Phi-\pi$, respectively.
As is shown in Appendix A, the scaling laws for even and odd $N$ may be
different. However, making only the flux transformation of 
Eq.\ (\ref{eq:pc-symmetry})
allows us to obtain good asymptotic results from a single fit to the ensemble 
of data points for even and odd $N$. A second-order polynomial fit describes 
very well the deviation of the logarithm of the persistent 
current from its asymptotic value.

The results presented in Fig.~\ref{fig:pc-scaling} indicate that the symmetry 
(\ref{eq:pc-symmetry}) holds even in the presence of electron-electron 
interaction and is independent of the interaction strength $U$. 
This provides numerical evidence that it should be possible to 
relate the persistent current in the presence of an interacting region to the 
persistent current for a non-interacting scattering problem.

\begin{figure}[tb]
\centerline{\includegraphics[width=\figwidth]{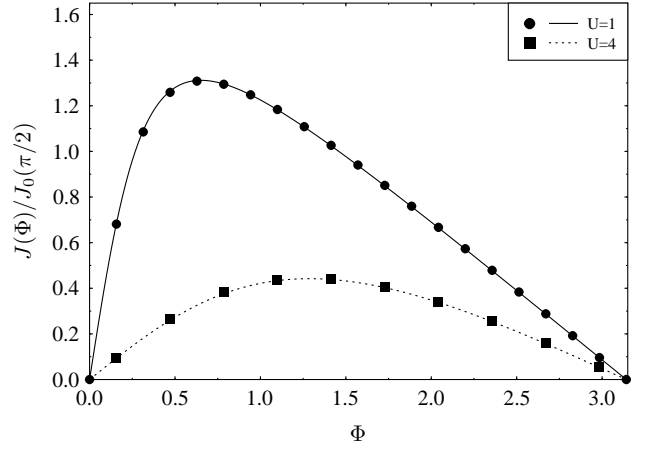}}
\vspace{2mm}
\caption[]{\label{fig:fluxdep} 
The flux dependence of the persistent current for a system size 
$L_\mathrm{S}=6$ and half filling is shown for interaction strengths $U=1$ and 
$4$. The points represent DMRG results extrapolated to the limit of infinite 
leads (see Fig.~\ref{fig:pc-scaling}). The lines represent the theoretical 
result (\ref{eq:jfluxeven}) for a ring with a non-interacting scatterer 
and transmission amplitudes $|t|=0.938$ (solid line) and $|t|=0.425$ (dotted
line).} 
\end{figure}

The flux dependence of the persistent current $J(\Phi)$ for an even number of 
particles, extrapolated to the limit of an infinite 
lead, is presented in Fig.~\ref{fig:fluxdep} for moderate and strong 
interaction, $U=1$ and $U=4$, respectively. At the filling factor 
$\nu=1/2$ used here, the interaction effects are expected to be most 
important. As can be seen from Fig.~\ref{fig:fluxdep}, the flux dependence of 
the persistent current is described very well by the expression 
(\ref{eq:jfluxeven}) for the non-interacting case with transmission 
amplitudes of $|t|=0.938$ (solid line) and $0.425$ (dotted line) for $U=1$ 
and $4$, respectively.

This demonstrates that, in the limit $L\to \infty$, the 
zero-temperature persistent current of a ring containing an 
interacting region is quantitatively described by the persistent 
current of a ring with a scatterering region. A single parameter, the 
interaction-dependent elastic transmission coefficient at the Fermi energy 
$\left|t(E_{\rm F},U)\right|^2$ suffices to characterize the interacting 
sample, at least as far as the flux dependence of the ground state energy
at zero temperature is concerned.

We emphasize that the DMRG technique employed here to calculate the 
persistent current of the ground state of the Hamiltonian (\ref{eq:hamil}) 
does not rely on any assumption. In particular, the DMRG technique does not 
require that the correlated nanosystem must be a Fermi liquid. But the 
fact that the expressions (\ref{eq:jfluxodd}) and (\ref{eq:jfluxeven}) for 
the persistent current hold in the infinite lead length limit even in the 
presence of an interacting region provides strong evidence that the Fermi 
liquid behavior is retained in this limit. This result is in agreement with 
the theoretical expectation mentioned above.

Our findings constitute a numerical ``proof'' that the extension of the 
relation between persistent current and transmission from a non-interacting to 
an interacting system is correct. Assuming that the composed ring forms a Fermi 
liquid, a discussion of the relation between the persistent current and the
conductance had already been given in \cite{rejec03a}. Together with the results
of this section, this opens a road towards the calculation of the 
conductance for interacting nanosystems. 

\section{Conductance from transmission for interacting scatterers}
\label{sec:condfromtrans}

Instead of the persistent current, we will, in the following,
mostly work with the charge stiffness defined as:
\begin{equation}\label{eq:stiffness}
D=(-1)^N \frac{L}{2}\big( E(0)-E(\pi)\big) 
\end{equation}
which describes the change of the ground-state energy from periodic to 
antiperiodic boundary conditions. The factor $(-1)^N$ renders $D$ positive
because the many-body ground state is diamagnetic for odd $N$ while it is 
paramagnetic for even $N$. This fact was proven by Leggett \cite{Leggett} 
for spinless fermions in the presence of arbitrary one-body potentials and 
arbitrary strength of electron-electron interactions. We prefer to
work with the charge stiffness
$D$ instead of the persistent current $J$ because it allows to avoid the use
of a complex implementation of the DMRG algorithm and thus reduces the 
numerical effort. 

For the case of a non-interacting scatterer, the flux dependence of the 
ground-state energy is derived in Appendix \ref{sec:noninteracting}.
>From Eqs.\ (\ref{eq:stiffleadingapp}) and (\ref{eq:e0even1}) it follows 
that for the limit of infinite lead length we have
\begin{equation}\label{eq:continuous}
D=\frac{\hbar v_\mathrm{F}}{2}
\left[\frac{\pi}{2}-\mathrm{Arccos}(|t(k_F)|)\right] \, ,
\end{equation}
independent of the parity of $N$.
Solving (\ref{eq:continuous}) for the tunneling probability 
yields \cite{molin03}
\begin{equation}
\left|t(k_\mathrm{F})\right|=\sin\left( \frac{\pi}{2}\frac{D}{D^0}\right) \ ,
\label{eq:stiff}
\end{equation}
where $D^0$ is the charge stiffness for a clean ring of length
$L$ in the absence of electron-electron interactions. We note that for
weak transmission ($|t| \ll 1$), $D$ is proportional to $|t|$.

We have verified that the transmission coefficients calculated from the
stiffness using Eq.\ (\ref{eq:stiff}) as
described in Ref.\ \cite{molin03} coincide with the ones
obtained by fitting the full flux dependence of the persistent current 
(Fig.~\ref{fig:fluxdep}) to a precision better than $0.5 \%$. 

\subsection{Scaling of the stiffness and extrapolation to infinite
  lead length}
\label{sec:scaling-easy}

As already discussed in Section~\ref{sec:flux-dependence}, the limit of an
infinitely long lead is required in order to obtain the conductance. While
for the persistent current, we had been restricted to rather small ring sizes,
the charge stiffness allows us to numerically treat rings almost an order of
magnitude larger. This will enable us to take a closer look at the scaling of
the charge stiffness with $1/L$, even in difficult cases like in 
the presence of transmission resonances.

\begin{figure}[tb]
\centerline{\includegraphics[width=\figwidth]{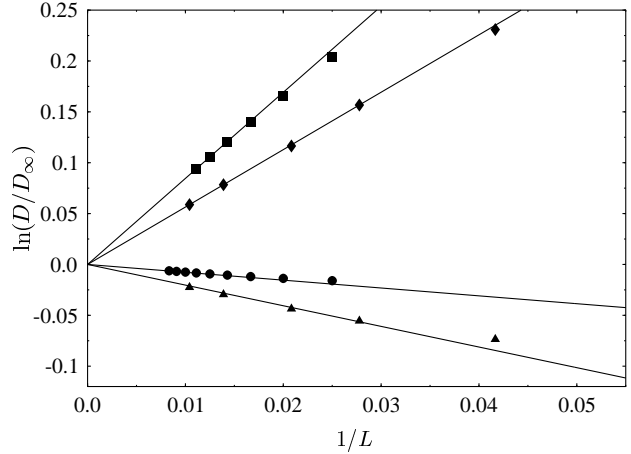}}
\vspace{2mm}
\caption[]{\label{fig:scaling-easy} 
The scaling of the logarithm of the charge stiffness with the ring
size is shown for systems at half filling and $L_\mathrm{S}=20, U=3$ 
(squares), $L_\mathrm{S}=12, U=4$ (diamonds), $L_\mathrm{S}=17, U=1$ 
(circles), and $L_\mathrm{S}=13, U=2$ (triangles). The lines are 
linear fits to the large-$L$ behavior, providing the extrapolation to 
infinite ring size, \textit{i.e.} $1/L\to 0$.}
\end{figure}

As the derivation of the charge stiffness as a function of the 
transmission amplitude in appendix \ref{sec:noninteracting} shows, 
the charge stiffness for large rings can be expanded in powers of $1/L$. 
In the limit $L\to\infty$, a non-vanishing contribution given by 
(\ref{eq:continuous}) allows us to determine the conductance. 

The leading corrections (\ref{corrstiffodd}) and (\ref{corrstiffeven}) 
for odd and even number of particles, respectively, are of order $1/L$. 
Essential for the relevance of these corrections is their dependence
on the derivatives with respect to $k$ of the transmission $|t|$ 
and the relative phase shift $\delta\alpha$ characterizing the scattering 
region. $\mathrm{d}\delta\alpha/\mathrm{d}k$ is proportional to the Wigner 
delay time \cite{Wigner}. At resonances, the two derivatives may become very 
large. Then, only rings of circumference $L \gg \mathrm{d}|t|/\mathrm{d}k,
\mathrm{d}\delta\alpha/\mathrm{d}k$ allow to perform a reliable extrapolation 
to the asymptotic limit. This situation will be discussed in 
Section~\ref{sec:scaling-difficult}. Outside resonances, we found 
that the extrapolation can usually be performed with rings about three or 
four times as large as the scattering region.

In Fig.~\ref{fig:scaling-easy}, we present the deviation of the logarithm
of the charge stiffness from its asymptotic value as a function of the inverse
circumference $L$ of the ring. This plot is the analogue of 
Fig.~\ref{fig:pc-scaling} where the scaling of the persistent current was
depicted, but now the size of the nanosystem is up to a factor of 
three larger. In all cases shown here, we are far away from any
resonance.

The scaling with the ring size has been described by different laws in the 
literature. A parabolic fit was assumed in the first paper of Favand and 
Mila \cite{favan98} while a linear fit to the deviations of the logarithm 
was employed in our previous paper \cite{molin03}. Different polynomial 
scalings were compared by Meden and Schollw{\"o}ck \cite{meden03}. In the
present work, we have used a linear scaling for the deviations of 
$\ln(D)$. A second-order fit becomes necessary when numerical limitations 
prevent us from attaining sufficiently large ring sizes as it has been
the case for the persistent current (cf.\ Fig.~\ref{fig:pc-scaling}). 

For the extrapolation of the charge stiffness in the cases presented 
in Fig.~\ref{fig:scaling-easy}, it is sufficient to use the scaling law 
\begin{equation}\label{eq:scaling}
D(U,L_\mathrm{S},L) = D_{\infty}(U,L_\mathrm{S})\exp{\left(
                                  \frac{C(U,L_\mathrm{S})}{L}\right)} 
\end{equation} 
to determine the asymptotic value $D_{\infty}(U,L_\mathrm{S})$. The 
conductance is then obtained from (\ref{eq:stiff}) as 
\begin{equation}
g=\sin^2 \left(\frac{\pi}{2}\frac{D_{\infty}}{D^0}\right)\,.
\end{equation}
This procedure had been used in Ref.\ \cite{molin03} to compute the 
influence of the interaction strength on the conductance of correlated 
nanosystems at half filling. The conductance of a clean system decreases with
the interaction strength (see the solid line in Fig.~\ref{fil8}) 
for even numbers of particles, and remains perfect ($g=1$) for odd numbers 
of particles independently of the interaction strength. 

\subsection{Scaling close to transmission resonances}
\label{sec:scaling-difficult}

\begin{figure}[tb]
\centerline{\includegraphics[width=0.7\figwidth]{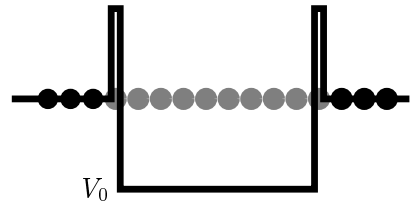}}
\vspace{2mm}
\caption[]{\label{fig:double} Sketch of the site potentials used for the 
double barrier system. Electron-electron interaction is present only on the
grey sites.}
\end{figure}

The leading correction (\ref{corrstiffodd}) or (\ref{corrstiffeven})
to the charge stiffness may play an important role close to transmission 
resonances, where the Wigner delay time and $\mathrm{d}|t|/\mathrm{d}k$ 
are large. We illustrate the difficulties in the extrapolation procedure
present in this case by considering a nanosystem separated from the auxiliary 
lead by two tunnel barriers (cf.\ Fig.~\ref{fig:double}). In order to
tune the Fermi energy of the ring to a resonance, we introduce an 
electrostatic potential $V_0$ between the tunnel barriers of height 
$V_\mathrm{b}=1$. A single-particle term $V_\mathrm{b}(n_{1}+n_{L_\mathrm{S}})+
V_0\sum_{i=2}^{L_\mathrm{S}-1}n_i$ is thus added to the 
Hamiltonian (\ref{eq:hamil}). The electron-electron interaction is present 
on all $L_\mathrm{S}$ sites including the two barrier sites but the lead 
remains non-interacting as usual. We note that the additional potential
$V_0$ will change the electron density in the nanosystem. 

\begin{figure}[tb]
\centerline{\includegraphics[width=\figwidth]{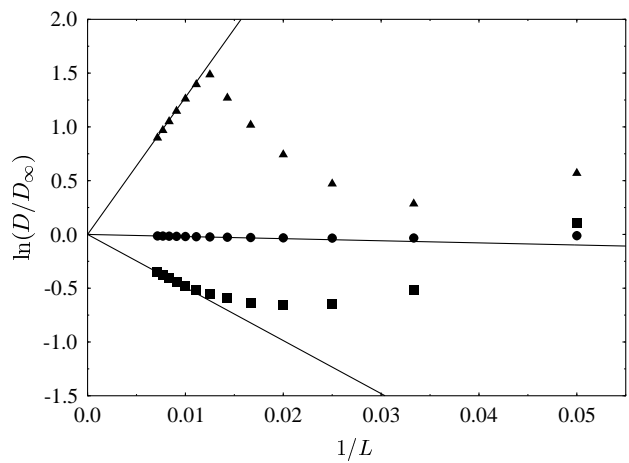}}
\vspace{2mm}
\caption[]{\label{scalwell} Scaling towards the asymptotic value of the 
stiffness $D$ for a weakly coupled nanosystem with $L_\mathrm{S}=10$ and
$U=1$. The circles, squares, and triangles correspond to electrostatic 
potentials $V_0=-0.8$ (out of resonance), $V_0=-1.4$ (just to the right 
of a resonance), and $V_0=-1.5$ (just to the left of a resonance), 
respectively.}
\end{figure}

Resonances occur whenever the ground state energies of the ring with $N+1$ 
particles and $N$ particles inside the double-barrier system are identical. 
In the absence of electron-electron interaction, this implies that the energy 
of the first unoccupied single-particle state of the well between the barriers 
lines up with the Fermi energy of the leads. When the degeneracy between 
ground states with different number of particles in the system appears, 
the energetic cost for transporting a particle through the system is zero 
and the transmission is one. 

For the reasons discussed in the previous section
(see also Appendix \ref{sec:noninteracting}), 
this case is characterized by a slow convergence towards the  
limit $L_\mathrm{L}\to\infty$. 
Large lead lengths are then needed because the very
rapid changes of the transmission as a function of $k$ lead to 
large corrections. Another reason consists in the difficulty 
to maintain the resonance condition for the
electron density of the nanosystem in the scaling procedure.
However, even in this unfavorable case, the conductance can be obtained by 
going to larger systems and taking the asymptotic value with a greater 
care than for the non-resonant case.  

Fig.~\ref{scalwell} shows for the example of a double-barrier system
how one can extrapolate to the asymptotic value of the stiffness in 
three cases, one favorable and two unfavorable. The ratio $\ln(D/D_{\infty})$ 
is given as a function of the inverse total length of the ring. 
The circles correspond to $V_0=-0.8$ and $U=1$, situated in the valley 
between two resonances where the conductance is small. In this case, the 
extrapolation is straightforward and the slope is very small. The other two
cases are different. Taking $V_0=-1.4$ and $U=1$ (depicted by squares), we 
are just to the right of a resonance. The corrections to the scaling 
formula (\ref{eq:scaling}) are very large for small ring sizes, and a naive 
extrapolation from there can give wrong values (even $g > 1$) for 
the conductance. In order to test that the asymptotic value for $D$ is 
approached, one calculates the parameters $C$ and $D_{\infty}$ of the scaling 
formula (\ref{eq:scaling}) for two different values of $L$ and one continues 
to increase $L$ until the slope $C$ and the asymptotic stiffness $D_{\infty}$ 
converge to constant values. 
In the case $V_0=-1.5$ shown by triangles in  Fig.~\ref{scalwell}, 
we have first determined $C$ and $D_{\infty}$ assuming the scaling law 
(\ref{eq:scaling}) for $L=30$ and $L=40$. Because the procedure gives
different results when we take $L=40$ and $L=50$, we were forced to  
increase $L$. Since the values for $C$ and 
$D_{\infty}$ obtained with $L=120$, 130 and 140 do not vary, we assume 
that one has reached the asymptotic regime. This procedure can require 
large values of the total length $L$ of the ring, which are difficult 
to reach for large filling factors $\nu$. Using a fit with more parameters
can be an option when the convergence is slow, but the extrapolation
must be done very carefully. The behavior of the stiffness $D$ as a 
function of the length $L$ in this last example is quite complicated 
because the density in the lead cannot be kept perfectly uniform and
therefore the resonances move as a function of the increasing size of 
the ring. This extreme case illustrates the potential difficulties
which must be solved in order to get reliable values for $g$ in the 
vicinity of transmission resonances from this method. For $V_0=-1.5$ 
(just to the left of a resonance), the slope has changed sign and 
we still need to go to big ring sizes for a reliable extrapolation.

\begin{figure}[tb]
\centerline{\includegraphics[width=\figwidth]{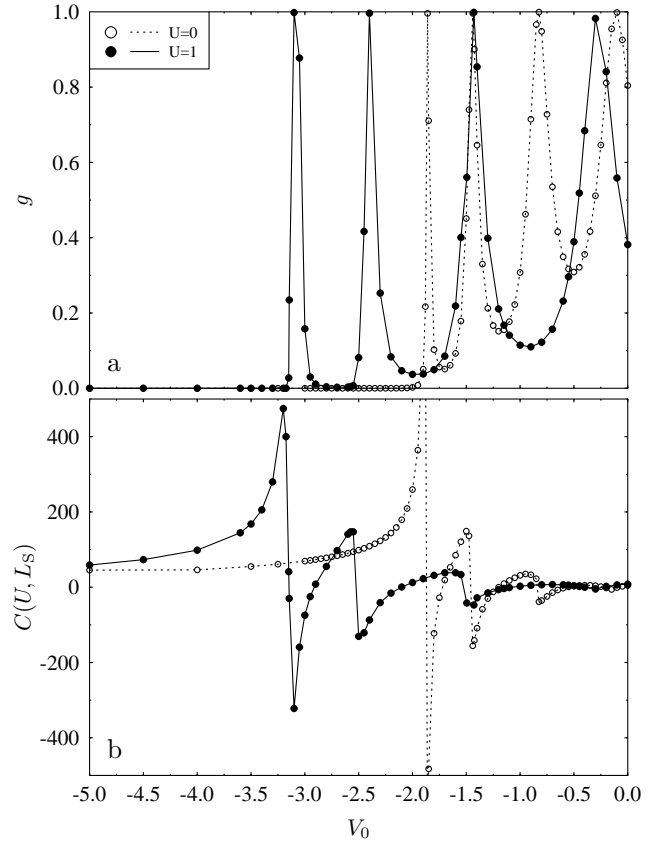}}
\vspace{2mm}
\caption[]{\label{fig:well} (a) Conductance $g$ and (b) slope 
$C(U,L_\mathrm{S})$ of the scaling law (\ref{eq:scaling}) are shown as a 
function of the electrostatic  well potential $V_0$ for the configuration of 
Fig.~\ref{fig:double}. Results for interaction strength $U=1$ are indicated by
full symbols and a solid line while the open symbols and the dotted line
represent results for the non-interacting case ($U=0$).}
\end{figure}

In Fig.~\ref{fig:well}a we depict the results of the conductance,
evaluated using the previous extrapolations for the two-barrier system.
We compare the results for $U=0$ and $U=1$. The values for $U=0$ have 
been obtained in the same way as the values for $U=1$, using DMRG and 
the extrapolation. They are found to agree with results from a 
non-interacting Green function calculation. The fact that we do obtain 
perfect conductance ($g=1$) at resonances supports our claim that the
asymptotic procedure is capable of yielding the correct
transport properties. In Fig.~\ref{fig:well}b we show the 
slope $C(U,L_\mathrm{S})$ of the scaling law (\ref{eq:scaling}). 
As one can see, the resonance structure is clearly reflected by the slope 
of the scaling curves.
The jumps in the slope coincide with the values for which the 
dimensionless conductance approaches its maximum value of one. The slope is 
closely related to the behavior of $\mathrm{d}|t|/\mathrm{d}k$.
As expected, the interaction $U$ changes the
position of the peaks and their widths.

\section{Conductance outside half filling}
\label{sec:arbitrary-filling}

As stated in the introduction, most of the applications of the
embedding method have so far been restricted to half filling. In the
previous section, we have maintained half filling in the average over
the composed ring, but the filling of the correlated system itself
depended on the potential $V_0$ between the barriers. As an even more
general situation, we now consider the case where the filling in the
composed ring has an arbitrary value $\nu$. 
The half-filled systems exhibit particle-hole symmetry,
and therefore the compensating potential $V_+=1/2$ required to yield 
charge neutrality inside the nanosystem is known \textit{a priori}. If we 
want to ensure a given constant filling $\nu$ for the nanosystem and
the lead even when the lead length is changed, $V_+$ becomes a
function of the interaction strength
and the ring size. In this section, we extend the method to nanosystems outside
half filling which are well coupled to the lead. By choosing the appropriate
particle number, the same filling is imposed in the auxiliary lead in order
to obtain the transmission coefficient $\left|t(E_{\rm F},U)\right|^2$ 
at the corresponding Fermi energy and to ensure a better convergence towards 
the limit of infinite lead length. 

In order to determine $V_+$ for an arbitrary filling $\nu$, 
we begin with an initial guess for $V_+$ and calculate numerically the 
corresponding number of particles contained inside the nanosystem. 
Then, we adjust $V_+$ performing an iterative solution 
of the problem using the Newton-Raphston method. 
In principle, $V_+$ depends on $U$ and $L_\mathrm{L}$. For example, 
for $L_\mathrm{S}=8$, $\nu=3/8$ and $U=3$, $V_+$ varies from 0.1924 to 
0.1939 as $L_\mathrm{L}$ is doubled from 24 to 48. At a fixed interaction
strength, the dependence of $V_+$ on $L_\mathrm{L}$ becomes negligible 
beyond a certain $L_\mathrm{L}$, and can then be ignored. Therefore, the
iterative procedure has only to be performed until a limiting value for
$V_+$ has been attained. Then this value can be kept for larger ring
sizes from which $C$ and $D_{\infty}$ are determined, using the same 
scaling law as at half filling. 

\begin{figure}[tb]
\centerline{\includegraphics[width=\figwidth]{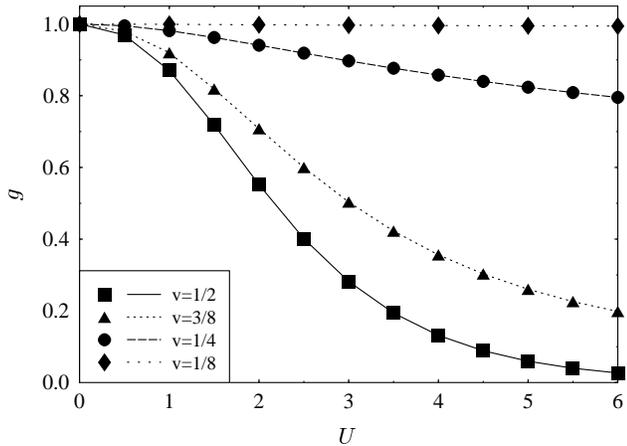}}
\vspace{2mm}
\caption[]{\label{fil8} 
Conductance as a function of $U$ for different filling factors $\nu$ of 
a correlated system of size $L_\mathrm{S}=8$.}
\end{figure}

In Fig.~\ref{fil8}, the conductance of a nanosystem of length 
$L_\mathrm{S}=8$ perfectly coupled to the lead 
is given as a function of the interaction strength $U$ at 
different filling factors $\nu$. Since the filling is kept uniform 
everywhere in the ring, the curves characterize 
$g(E_{\rm F},U)$ at the corresponding Fermi energy $E_{\rm F}$. 
At $\nu=1/8$ (short dashed line), in average only one particle 
is left in the nanosystem. In the absence of other particles to 
interact with, the
dimensionless conductance therefore equals one, independently of the
interaction strength. For larger filling factors, the conductance $g$
decreases with increasing interaction strength $U$ and this decay
becomes more pronounced as the filling factor is increased. The rather 
sharp drop of the conductance occurring at half filling around $U=2$ is a 
precursor of the Mott transition expected in the thermodynamic 
limit. The conductance above half filling can be obtained from 
$g(\nu)=g(1-\nu)$ as a consequence of particle-hole symmetry. The
influence of the interaction strength on the conductance is thus the
strongest at $\nu=1/2$ as expected.

\section{Conductance for disordered nanosystems}\label{sec:disorder}

Having demonstrated that the conductance of a correlated nanosystem
can be obtained from the charge stiffness after embedding it into a large
noninteracting ring, we now apply this method to the problem of interacting 
electrons in disordered systems. The effect of repulsive interactions in a 
disordered system is a controversial issue \cite{Pichard}.
It is often believed that interactions impede transport. This 
belief comes from perturbative arguments showing that interactions 
reduce the density of states at the Fermi level of a disordered metal 
\cite{altshuler} and open a gap for a strongly disordered 
insulator \cite{efros}. On the other hand, in the strong disorder limit 
zero temperature transport can be enhanced by an interaction-induced 
delocalization of the many-body ground state. This was demonstrated for the 
special case of half filling in Ref.\ \cite{molin03}. In the following,
we will study the role of the filling factor in the delocalization process. 

\begin{figure}[tb]
\centerline{\includegraphics[width=\figwidth]{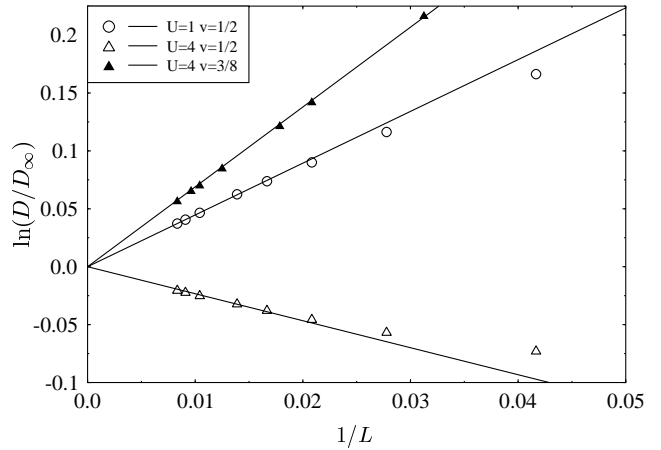}}
\vspace{2mm}
\caption[]{\label{scaldis} Scaling towards the asymptotic value of the 
stiffness $D$ for disordered samples ($W=5$). 
For the same disorder realization, two values of the interaction
are shown for half filling. Open triangles represent $U=4$ and 
open circles the case $U=1$. 
The negative slope in the former case corresponds to a charge 
reorganization and the conductance 
$g=0.34$ for $U=4$ is greater as compared to $g=0.018$ for $U=1$. Results
for the same disorder configuration with $U=4$ and $\nu=3/8$ are 
displayed with filled triangles. The conductance in this 
case is $g=0.0011$, demonstrating that the charge reorganization depends 
on the filling.}
\end{figure}

We include the disorder potential into the Hamiltonian (\ref{eq:hamil}) 
by adding a term 
\begin{equation}
H_\mathrm{dis}=W\sum_{i=1}^{L_\mathrm{S}}v_i n_i\,,
\label{eq:disorderterm}
\end{equation}
where $W$ denotes the disorder strength, and the $v_i$ are
independent random variables, equally distributed within the interval 
$[-1/2,1/2]$. The disorder potential is only present within the nanosystem
of length $L_\mathrm{S}$.

We start by verifying that the scaling towards infinitely large rings
also works in the presence of disorder. Fig.~\ref{scaldis} depicts the
dependence of the logarithm of the charge stiffness $D$ on the ring size
$L$ for a sample with $W=5$ for interaction strengths $U=1$ (circles) 
and $4$ (triangles). The disorder realization is the same in both cases.
The open symbols refer to $\nu=1/2$ while the full symbols correspond to
$\nu=3/8$. In all cases the scaling works well, and thus reliable values
for the conductance can be extracted.

The analysis of individual samples helps us to understand the physical
mechanisms involved when disorder and interactions are both relevant
\cite{smit1,smit2}. Studying the evolution of the ground state energy
or the electron density as a function of $U$, we can detect charge 
reorganizations at critical values of the interaction strength. For
the sample shown in Fig.~\ref{scaldis} we have, at half filling, 
a charge reorganization in the ground state structure around $U=4$. 
Charge reorganizations appear when a ground state configuration which
is well adapted to the non-interacting case, where the fermions are 
located in the minima of the disorder potential, changes towards a 
Wigner-like crystalline structure which is energetically favorable 
at strong repulsive interaction.
This resonant situation increases the conductance at the particular 
(sample dependent) cross-over value of the interaction. In other samples the 
charge reorganizations can occur at different values of the interaction
or can even be absent, depending on the disorder realization. 
Reducing the filling makes the charge reorganizations less
likely.  For the charge reorganizations of the disordered case,
we typically obtain a negative slope for the asymptotic scaling of $D$,
similar to the case of clean systems with odd number of particles 
\cite{molin03}. In both cases, a degeneracy of different charge 
configurations in the nanosystem occurs.

\begin{figure}[tb]
\centerline{\includegraphics[width=\figwidth]{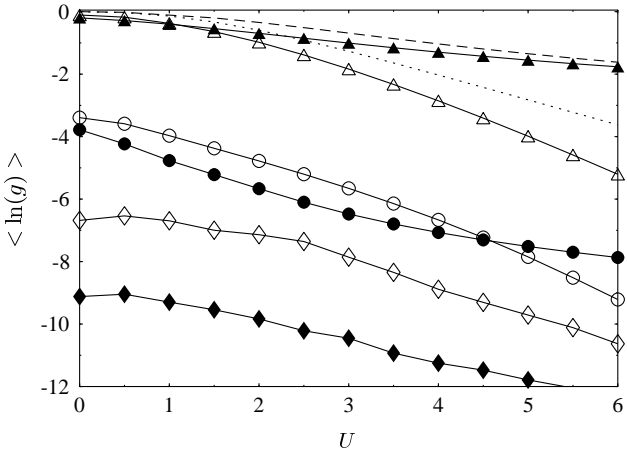}}
\vspace{2mm}
\caption[]{\label{disor} Logarithmic ensemble average of the conductance as a 
function of the nearest neighbor repulsion $U$ for a disordered nanosystem 
of length $L_{\rm S}=8$. The open symbols correspond to half-filled 
nanosystems, and the filled symbols to a filling factor $\nu=3/8$. The 
triangles correspond to $W=1$, the circles to $W=5$ and the diamonds to 
$W=9$. For $W=0$, the dashed line corresponds to $\nu=3/8$ and 
the dotted line to $\nu=1/2$, respectively.}
\end{figure}

In Fig.~\ref{disor}, the ensemble average of the logarithm of $g$ 
is given as a function of $U$ for disorder strengths $W=1$ (triangles),
$W=5$ (circles), and $W=9$ (diamonds) and filling factors $\nu=1/2$ (open 
symbols) and $\nu=3/8$ (full symbols). One can see from the
increase of the average conductance at weak interaction in the strongly 
disordered case, $W=9$, that the nearest neighbor interaction has stronger 
delocalization effects around half filling.
The results for the clean case show Mott insulator like behavior at half 
filling (dotted line). 
The decay of the typical value of $g$ as a function of $U$ is faster for 
the half-filled case than for $\nu=3/8$. When we introduce a
random potential in the nanosystem,  
the reduction of the typical conductance due to localization effects is 
more important outside half filling. 

The larger the density, the better 
is the screening of the random potentials. In the case of weak disorder,
$W=1$, this gives rise to a crossing of the curves with $\nu=1/2$ and 
$\nu=3/8$ as $U$ increases. For stronger disorder, this crossing occurs at 
larger values of $U$ ($U=4.5$ for $W=5$) and for very strong disorder ($W=9$) 
the crossing cannot be observed in the figure. 

We can also see in Fig.~\ref{disor} that in the strong disorder case 
(here $W=9$), 
nearest neighbor interactions can favor transport. This enhancement of the 
typical elastic transmission, and hence of the zero temperature conductance, 
is maximal around $U=0.5$ and, though mainly characteristic for half filling, 
it persists outside $\nu=1/2$.

The charge reorganization induced by repulsive interactions in 
strongly disordered systems and its associated delocalization effect
was first observed in the persistent current of nanosystems 
\cite{smit1,smit2} forming a ring (without the auxiliary lead introduced
within the embedding approach). As our results demonstrate, the
same effects can be found in the conductance $g$.
Considering a given nanosystem, one observes a similar resonance 
structure \cite{molin03} as for the persistent current \cite{smit1,smit2}, 
although the individual peaks are wider for the conductance than for
the persistent current.

\section{Summary}\label{sec:summary}

The residual conductance of a correlated nanosystem can be obtained from
the charge stiffness or from the persistent current of a ring composed of the
system and an auxiliary non-interacting lead. Using DMRG for spinless
fermions, we have numerically studied basic properties of this embedding
approach. In particular, we have demonstrated that the flux dependence of the
persistent current for an interacting system and a non-interacting lead agrees 
with the flux dependence of a non-interacting ring with a scatterer, in the 
limit of infinite lead length. This allows to extract the interaction dependent 
transmission coefficient of the interacting system, and hence its residual 
conductance.

A detailed analysis of the finite-size corrections has been performed
for the charge stiffness. The main features of these corrections can
be understood from the analysis of the non-interacting case. Away from 
transmission resonances, we obtain a very good scaling behavior 
already for not too large lead lengths, and the conductance of the correlated 
nanostructure can be readily obtained. Close to resonances the asymptotic 
limit of large lead lengths is problematic and only by considering very 
long leads we obtain the correct asymptotic behavior. Even in these special 
cases, the results for the conductance agree with our expectation for the
resonant tunneling behavior in a double barrier structure.

It is straightforward only at half filling to keep the electron density
in the correlated system fixed when changing the ring size. We have
demonstrated that an extension of the method to arbitrary filling factors
is feasible provided the compensating potential is adjusted appropriately.
For clean samples, it was observed that the decrease of the conductance
with increasing interaction strength is strongest at half filling and
becomes weaker as the filling factor changes towards smaller or larger
values.

Another extension consists in the introduction of disorder in the correlated
system.  Charge reorganizations of the ground state
appear at sample-dependent values of the interaction strength, affecting
the long lead scaling and the asymptotic values. In the ensemble averages, 
we obtain for weak disorder a decreasing conductance as a function
of the interaction strength. However, for strong disorder we
have shown that a nearest neighbor repulsion can enhance the 
average of the logarithm of the conductance for spinless fermions in 
one-dimensional samples. This enhancement persists outside half filling, 
although it becomes weaker. 

So far, the approach is still limited to spinless fermions and single-channel 
leads, although the system itself can be arbitrary. Nevertheless, the method
is well suited to study the role of the contacts between the nanosystem and the 
leads. Furthermore, interesting phenomena like even-odd oscillations of the 
conductance with the number of fermions were found with this approach 
\cite{molin03,molin04}. In the absence of spin-flip scattering,
the generalization to electrons with spin is straightforward. Indeed, first
calculations for the Hubbard model have already been performed \cite{molin03}. 
These and further issues will be explored in more detail in future work.

\begin{acknowledgement}
RAM wishes to thank J.\ S\'egala for reminding him of some properties
of the Chebyshev polynomials. 
We gratefully acknowledge financial support from the European Union 
through the RTN program (Contract No.\ HPRN-CT-2000-00144).
PS  was supported by the Center for
Functional Nano\-struc\-tures of the Deutsche Forschungsgemeinschaft
within project B2.
\end{acknowledgement}

\appendix

\section{Flux dependence of the ground state energy for large rings
  with a small non-interacting scattering region} 
\label{sec:noninteracting}
 
In this appendix, we discuss the flux dependence of the ground state energy
for a ring containing a non-interacting scatterer. 
The scattering region of length $L_\mathrm{S}$ is connected to a 
disorder-free lead of length $L_\mathrm{L}$. This arrangement is
closed to a ring of total length $L=L_\mathrm{S}+ L_\mathrm{L}$, as shown in 
Fig.~\ref{fig:system}. We present a systematic expansion in powers of 
$1/L$ starting from the limit of infinite lead length for the 
flux-dependent part of the ground state energy. This leads to 
analytic expressions for the asymptotic values of the persistent current 
and the charge stiffness, as in Ref.\ \cite{gogolin}. 
We extend this theory by calculating the first finite-size corrections to 
the flux-dependent part of the energy and the charge stiffness.
These corrections are important to
understand the way in which the asymptotic values are approached when
we extrapolate to infinite ring size.

The one-particle eigenenergies of the ring are given 
by the quantization condition
\begin{equation}
\mathrm{det}\left(I-M_\mathrm{L} M_\mathrm{S}\right)=0 \ ,
\label{quantcont}
\end{equation}
where $M_\mathrm{S}$ and $M_\mathrm{L}$ are the 
transfer matrices of the system and the lead, respectively. 
In the presence of time-reversal symmetry, the transfer matrix of a 
one-dimensional scatterer can be expressed in terms of three 
independent angles $\alpha$, $\theta$ and $\varphi$:
\begin{equation}
\begin{aligned}
M_\mathrm{S} &= \left(\begin{array}{cc}
1/t^* & r^*/t^*\\
r/t & 1/t
\end{array}\right) \\ 
&= \frac{1}{\sin{\varphi}}
\left(\begin{array}{cc}
e^{i \alpha}/\sin{\theta} & -i\cot{\theta}+\cos{\varphi} \\
i\cot{\theta}+\cos{\varphi} & e^{-i \alpha}/\sin{\theta}
\end{array}\right)
\, , 
\end{aligned}
\label{eq:ms}
\end{equation}
where the two components correspond to right and left moving particles
while $r$ and $t$ are the reflection and transmission amplitudes,
respectively. The angle $\alpha$ is the 
phase-shift associated with the scattering region. Whenever the 
right-left symmetry is respected, we can set $\varphi=\pi/2$, 
and the expression of 
$M_\mathrm{S}$ simplifies considerably. However, this symmetry 
requirement is not satisfied for disordered samples. 
In the general case the 
transmission amplitude is given by 
$t= e^{\mathrm{i} \alpha} \sin{\theta} \sin{\varphi}$. 

The transfer matrix of a lead of length $L_\mathrm{L}$ for a state 
with wave number $k\ge0$ reads
\begin{equation}
M_\mathrm{L} = \exp{(\mathrm{i} \Phi)} \left(\begin{array}{cc}
\exp(\mathrm{i}kL_\mathrm{L}) & 0\\
0 & \exp(-\mathrm{i}kL_\mathrm{L})
\end{array}\right).
\label{eq:ml}
\end{equation}
Here, we have made use of the fact that the flux can be transformed into
a boundary condition which may be prescribed in the lead.

Inserting the transfer matrices (\ref{eq:ms}) and (\ref{eq:ml}), the eigenvalue
condition (\ref{quantcont}) yields
\begin{equation}
\cos(\Phi)=\frac{1}{\left| t(k) \right|}\cos\big( kL+\delta \alpha (k) 
\big) \ .
\label{eq:quantcont2}
\end{equation}
Here, we have introduced the phase shift 
$\delta \alpha = \alpha - k L_\mathrm{S}$ of the scattering region relative
to a perfect lead of the same length $L_\mathrm{S}$. The solution of 
(\ref{eq:quantcont2}) yields the quantized momenta $k$ of the energy 
eigenstates in the lead.

Since both, $t$ and $\delta \alpha$ are functions of $k$, it is in 
general impossible to obtain an analytic solution of (\ref{eq:quantcont2}). 
However, progress can be made in the asymptotic limit of large $L$, which 
was worked out by Gogolin and Prokof'ev \cite{gogolin} in their study of 
the persistent current. We extend their approach to calculate the first
finite-size corrections of the charge stiffness. Furthermore, a generalization 
to arbitrary dispersion relation in the lead allows us to discuss continuum 
and tight-binding models at the same time.

The eigenvalue condition (\ref{eq:quantcont2}) can be rewritten as
\begin{equation}
k = k^0_n + \frac{1}{L} f_{\pm}(k,\Phi) \ .
\label{eq:eveq}
\end{equation}
Here, $k^0_n=2\pi n/L$ with $n\ge 0$ denotes the eigenvalues in the case of 
perfect transmission with $\vert t\vert=1$ and $\delta\alpha=0$. Following the 
notation of Ref.\ \cite{gogolin}, we have furthermore introduced
\begin{equation}
f_{\pm}(k,\Phi)= \pm \mathrm{Arccos} \left( \left|t(k) \right|
\cos \Phi \right) - \delta\alpha(k) \  .
\end{equation}
By Arccos, we denote the principal branch of the inverse cosine function that 
takes values in the interval $[0,\pi]$. In order to ensure a positive value 
for $k$, $f_-(k,\Phi)$ should not be used for the case $n=0$. The splitting of
the solutions of (\ref{eq:eveq}) corresponding to ``+'' and ``-''
cannot exceed the spacing $2\pi/L$ between the $k_n^0$, provided that
$\delta\alpha(k)$ is smooth on this scale. This is the case in the
limit $L\to\infty$ and ensures that the order of the solutions with
respect to energy is given by $n$. 

Iterating (\ref{eq:eveq}) and expanding $f_{\pm}$ for large systems, 
we obtain the expansion 
\begin{equation}\label{eq:kn}
\begin{aligned}
k_n^{\pm}= & k^0_n+\frac{1}{L}f_{\pm}(k^0_n,\Phi)\\
&+\frac{1}{L^2}f_{\pm}(k^0_n,\Phi)
\left( \frac {\partial f_{\pm}(k,\Phi)} {\partial k} \right)_{k=k_n^0} \\
&+\frac{1}{2L^3}\frac{\partial}{\partial k}
\left(f^2_{\pm}(k,\Phi)
\frac {\partial f_{\pm}(k,\Phi)} {\partial k} \right)_{k=k_n^0} 
  +  O \left(\frac{1}{L^4}\right)
\end{aligned}
\end{equation}
for the solutions of (\ref{eq:eveq}) in powers of $1/L$. 
Such an expansion is problematic in the vicinity of resonances, 
when $\mathrm{d}\delta \alpha/\mathrm{d}k$ and  
$\mathrm{d}|t|/\mathrm{d}k$ are very large. Then, the expansion is 
valid only for sufficiently large $L$. 

We now calculate the ground state energy of the system as a function of the 
flux to order $1/L^2$. The dispersion relation in the perfect lead will be
denoted by $\epsilon(k)$. Using (\ref{eq:kn}), we start by expanding the 
one-particle energies in powers of $1/L$ and obtain
\begin{equation}
\begin{aligned}
\epsilon(k^\pm_n) =& \epsilon(k^0_n)+\frac{1}{L}
\left(\frac{\partial \epsilon}{\partial k} f_{\pm}(k,\Phi)\right)_{k=k_n^0}\\
&+\frac{1}{2L^2}\frac {\partial } {\partial k} 
\left(\frac{\partial \epsilon}{\partial k} f_{\pm}^2(k,\Phi)\right)_{k=k_n^0}
 \\
  &+ \frac{1}{6L^3}\frac{\partial^2}{\partial k^2} 
\left(\frac{\partial \epsilon}{\partial k}f_{\pm}^3(k,\Phi)\right)_{k=k_n^0}
+ O\left(\frac{1}{L^4}\right)\, .
\label{eq:epsilon2}
\end{aligned}
\end{equation}

For an odd number $N$ of spinless electrons
in the ring, all occupied states $n$ come in pairs ([$n$,-] and 
[$n$,+]), except for the one corresponding to $n=0$. 
The total ground state energy then reads
\begin{equation}
\begin{aligned}
&E_0^\mathrm{odd}(\Phi)= \epsilon(k_0^+) 
+ \sum_{n=1}^{n_\mathrm{F}} [\epsilon(k_n^+) + \epsilon(k_n^-)] \\
&\;= \epsilon(0)
+\frac{1}{2L^2}\left(\frac{\partial^2\epsilon}{\partial k^2}
\left[\mathrm{Arccos}(|t|\cos\Phi)-\delta\alpha\right]^2\right)_{k=0}\\
&\quad + \sum_{n=1}^{n_\mathrm{F}} \left\{ 2 \epsilon(k_n^0) - 
\frac{2}{L} \left(\frac{\partial \epsilon}{\partial k}\right)_{k=k_n^0} 
\delta\alpha(k_n^0)\right. \\ 
&\quad+ \frac{1}{L^2}
\frac{\partial}{\partial k} \left(\frac{\partial\epsilon}{\partial k}
\left[\mathrm{Arccos}^2(|t|\cos\Phi)+\delta\alpha^2\right]\right)_{k=k_n^0}
\\ 
&\quad- \left. \frac{1}{3L^3}
\frac{\partial^2}{\partial k^2} \left(\frac{\partial\epsilon}{\partial k}
\left[3\delta\alpha\mathrm{Arccos}^2(|t|\cos\Phi)+\delta\alpha^3\right]\right)_{k=k_n^0}
\right\}\\ 
&\quad+ O \left(\frac{1}{L^3}\right)\, .
\label{genergy}
\end{aligned}
\end{equation}
The sum runs up to $n_\mathrm{F}=(N-1)/2$.
We have assumed $(\partial \epsilon/\partial k)_{k=0}=0$ and kept all terms 
which can give rise to contributions up to order $1/L^2$. 
The first term in the sum is the ground state energy in the absence of
scattering. For finite filling, \textit{i.e.} for $N$ of order $L$,
it is proportional to $L$ while the second term representing
the energy change due to the scattering potential is of order 1. 
The third and fourth terms are the leading flux-dependent corrections. 
Since we are interested in the persistent current and the charge stiffness,
these are the only terms in the sum which need to be considered further.  
Converting the sums over $n$ into integrals, these flux-dependent 
contributions can be expressed as 
\begin{equation}
\label{eq:int1}
\begin{split}
&\frac{1}{2 \pi L} \int\limits_{\pi/L}^{k_F+\pi/L}\mathrm{d}k
\frac{\partial}{\partial k} \left(\frac{\partial\epsilon}{\partial k}
\mathrm{Arccos}^2(|t|\cos\Phi)\right)\\
&\qquad= \frac {\hbar v_\mathrm{F}} {2 \pi L}\mathrm{Arccos}^2\big(|t(k_F)|
\cos(\Phi)\big) \\
&\qquad\quad+\frac{1}{2L^2}\left\{
\frac{\partial}{\partial k} \left(\frac{\partial\epsilon}{\partial k}
\mathrm{Arccos}^2(|t|\cos\Phi)\right)_{k=k_\mathrm{F}}\right. \\
&\qquad\quad\phantom{+\frac{1}{2L^2}\bigg\{ } 
-\left.\left(\frac{\partial^2\epsilon}{\partial k^2}
\mathrm{Arccos}^2(|t|\cos\Phi)\right)_{k=0}\right\}\\
&\qquad\quad+ O\left(\frac{1}{L^3}\right) 
\end{split}
\end{equation}
and 
\begin{equation}
\label{eq:int2}
\begin{split}
&-\frac{1}{2 \pi L^2} \int\limits_{\pi/L}^{k_F+\pi/L}\mathrm{d}k
\frac{\partial^2}{\partial k^2} \left(\frac{\partial\epsilon}{\partial k}
\delta\alpha \mathrm{Arccos}^2(|t|\cos\Phi)
\right)\\
&\qquad= -\frac{1} {2 \pi L^2}\left\{
\frac{\partial}{\partial k} \left(\frac{\partial\epsilon}{\partial k}
\delta\alpha \mathrm{Arccos}^2(|t|\cos\Phi)
\right)_{k=k_\mathrm{F}}\right.\\
&\qquad\quad\phantom{-\frac{1} {2 \pi L^2}\bigg\{}
-\left.\left(\frac{\partial^2\epsilon}{\partial k^2}
\delta\alpha \mathrm{Arccos}^2(|t|\cos\Phi)
\right)_{k=0}\right\} \\
&\qquad\quad+O\left(\frac{1}{L^3}\right)\, ,
\end{split}
\end{equation}
respectively. Here, $k_\mathrm{F}=2\pi n_\mathrm{F}/L$ is the Fermi wave 
number and $v_\mathrm{F}=(\partial\epsilon/\hbar\partial k)_{k=k_\mathrm{F}}$ 
is the Fermi velocity. Taking the derivative of the leading flux-dependent 
term of the ground state energy 
\begin{equation}
E_0^{\mathrm{odd}(1)}(\Phi)= 
\frac {\hbar v_\mathrm{F}} {2 \pi L}
\mathrm{Arccos}^2\big(|t(k_F)|\cos(\Phi)\big)
\end{equation}
with respect to the flux $\phi$, one obtains the asymptotic form of the 
persistent current given in (\ref{eq:jfluxodd}) for an odd number of particles. 
The leading order of the charge stiffness of (\ref{eq:continuous}) is 
obtained as
\begin{equation}\label{eq:stiffleadingapp}
\begin{aligned}
D^{(1)}&=
-\frac{L}{2}\left(E_0^{\mathrm{odd}(1)}(0)-E_0^{\mathrm{odd}(1)}(\pi)\right)\\
&= \frac{\hbar v_\mathrm{F}}{2} 
\left[\frac{\pi}{2}-\mathrm{Arccos}(|t(k_F)|)\right]\, .
\end{aligned}
\end{equation}   
As we will show below, this last result is independent of the parity
of the number of particles.

The first finite size-correction to these asymptotic values follows from 
the second-order contribution $E_0^{(2)}$ to the total energy. Using
the terms of order $1/L^2$ from (\ref{eq:int1}) and (\ref{eq:int2}), and
taking into account the contribution from the particle in the state $[0,+]$
in the second line of (\ref{genergy}), we obtain the correction to the charge
stiffness for an odd number $N$ of particles
\begin{equation}\label{corrstiffodd}
\begin{aligned}
D&^{\mathrm{odd}(2)}=-\frac{L}{2}
\left(E_0^{\mathrm{odd}(2)}(0)-E_0^{\mathrm{odd}(2)}(\pi)\right)\\
=& -\frac{1}{2L} \left\{\left(\left[\frac{\partial^2\epsilon}{\partial k^2}
(\delta\alpha-\pi)
+\hbar v_\mathrm{F}\frac{\mathrm{d}\delta \alpha}{\mathrm{d} k}\right] 
\left(\frac{\pi}{2}-\mathrm{Arccos}(|t|)\right)\right.\right.\\
&\phantom{ -\frac{1}{2L} \Bigg\{\Bigg(}
+\left.\hbar v_\mathrm{F}(\delta\alpha-\pi)\frac{\mathrm{d}|t|}{\mathrm{d}k} 
\frac{1}{\sqrt{1- |t|^2}}\right)_{k=k_\mathrm{F}}\\
&\phantom{ -\frac{1}{2L} \Bigg\{}+\left.\left(\frac{\partial^2\epsilon}{\partial k^2}
\delta\alpha \left(\frac{\pi}{2}-\mathrm{Arccos}(|t|)\right)\right)_{k=0}
\right\} \, .
\end{aligned}
\end{equation}
The last term vanishes if we assume that $|t(k=0)|=0$.

In order to treat also the case of an even number of particles, we 
subtract the contribution of the particle in the one-body state 
$[n_\mathrm{F},+]$ from the total energy of Eq.\ (\ref{genergy}) and obtain
\begin{equation}
\begin{aligned}
E_0^\mathrm{even}(\Phi)=&E_0^\mathrm{odd}(\Phi)-\epsilon(k^+_{n_\mathrm{F}})\\
=&E_0^\mathrm{odd}(\Phi)-\epsilon(k_\mathrm{F})
-\frac{1}{L}\left(\frac{\partial \epsilon}{\partial k} f_+(k,\Phi)\right)_{k=k_\textrm{F}}\\
&-\frac{1}{2L^2}\frac {\partial } {\partial k} 
\left(\frac{\partial \epsilon}{\partial k} 
f_+^2(k,\Phi)\right)_{k=k_\mathrm{F}}
+ O\left(\frac{1}{L^3}\right) \, .
\end{aligned}
\end{equation}

With these additional terms, one obtains the leading flux-dependent term of
the ground state energy for an even number of particles as
\begin{equation}\label{eq:e0even1}
E_0^{\mathrm{even}(1)}(\Phi)= 
\frac {\hbar v_\mathrm{F}} {2 \pi L}
\mathrm{Arccos}^2\big(|t(k_F)|\cos(\Phi-\pi)\big)\, .
\end{equation}
The derivative with respect to $\phi$ leads to the asymptotic form of the
persistent current of Eq.\ (\ref{eq:jfluxeven}) for an even number of
particles, and the result for the leading contribution to the charge 
stiffness agrees with (\ref{eq:stiffleadingapp}).  

For the first finite-size correction to the stiffness we obtain
\begin{equation}\label{corrstiffeven}
\begin{aligned}
D&^{\mathrm{even}(2)}=
\frac{L}{2}\left(E_0^{\mathrm{even}(2)}(0)-E_0^{\mathrm{even}(2)}(\pi)\right)\\
=& -\frac{1}{2L} \left\{\left[\frac{\partial^2\epsilon}{\partial k^2}
\delta\alpha
+\hbar v_\mathrm{F}\frac{\mathrm{d}\delta \alpha}{\mathrm{d} k}\right] 
\left(\frac{\pi}{2}-\mathrm{Arccos}(|t|)\right)\right.\\
&\phantom{ -\frac{1}{2L}\Bigg\{}
+\left.\hbar v_\mathrm{F}\delta\alpha\frac{\mathrm{d}|t|}{\mathrm{d}k} 
\frac{1}{\sqrt{1- |t|^2}}\right\}_{k=k_\mathrm{F}} \, ,
\end{aligned}
\end{equation}
which differs from the case of an odd number of particles 
(\ref{corrstiffodd}). 

>From Eqs.\ (\ref{corrstiffodd}) and (\ref{corrstiffeven}) one can
see that the $1/L$ scaling for approaching the asymptotic values of
the stiffness is problematic close to resonances, where 
$\mathrm{d}\delta \alpha/\mathrm{d} k$ and $\mathrm{d}|t|/\mathrm{d}k$
are large, and $|t|$ approaches 1. Assuming an isolated Breit-Wigner
resonance \cite{mucciolo}, the Wigner time is proportional to $g$ and
the corrections $D^{(2)}$ are essentially given by the half width of
the resonance. Outside resonances where $\delta\alpha\ll1$ and for small 
$|t|$, the leading
correction to the stiffness can be approximated by
\begin{equation}
D^{(2)} \approx -\frac{1}{2L}\hbar v_\mathrm{F}|t|
\frac{\mathrm{d}\delta \alpha}{\mathrm{d} k} \, .
\end{equation}
Therefore, one obtains for this case
\begin{equation}
\ln\left(\frac{D}{D_\infty}\right)\approx
\ln\left(\frac{D^{(1)}+D^{(2)}}{D^{(1)}}\right)
\simeq 
-\frac{1}{L}\frac{\mathrm{d}\delta \alpha}{\mathrm{d} k}
\, ,
\end{equation} 
and the Wigner time gives the slope of the scaling curve.
The above arguments are valid in the non-interacting case. However,
the intuition developed in this case is also useful to interpret our
numerical results for the interacting case.
 
\section{Conductance of a NSN region from persistent current}
\label{sec:super}

In this appendix we treat the case of a superconductor
between two metallic leads. This is a striking example of a correlated
system exhibiting non-Fermi liquid behavior. It will be demonstrated that the 
correct result for the conductance can be obtained from the persistent 
current by means of (\ref{eq:transpc}). 

\subsection{Double Andreev scattering}

The Andreev scattering at a NS junction, \textit{i.e.} the interface between 
a normal metal and a superconductor, is a well-known phenomenon. In an Andreev 
scattering process, an electron coming from the normal metal is reflected 
as a hole while a Cooper pair moves on in the superconductor. The linear 
conductance of the interface between the normal metal and the superconductor 
in the one-channel case is given by
\begin{equation}
\label{eq:Andreev}
G=\frac{4e^2}{h} \frac{T}{2-T}\,,
\end{equation}
where $T$ is the transmission probability in the normal metal 
\cite{Beenakker92,Lambert}. For the normal lead, $T=1$ and one 
gets that the resistance of a single normal-superconductor interface 
is the half of the resistance without interface. 

In the following, we will consider a NSN  double junction consisting of a 
clean superconducting layer of thickness $L_\mathrm{S}$ connected to 
normal-metal electrodes by perfect interfaces. It is assumed that the 
superconducting gap $\Delta(x)$ jumps at the interface from zero in the normal 
metal to  its full value $\Delta$ inside the superconductor 
\begin{equation}
\label{eq:gap} 
\Delta(x)=\Delta\Theta(x)\Theta(L_\mathrm{S}-x)\,,
\end{equation}
where $\Theta(x)$ is the step function. This approximation is common in 
the treatment of mesoscopic superconductors \cite{BeenakkerRMP}. 
Blonder, Tinkham and Klapwijk \cite{Blonder82} calculated the conductance
by solving the Bogoliubov-de Gennes equation with this rigid-boundary
condition and found for $T=1$ the linear conductance $G=2e^2/h$. This
result can be understood by taking two Andreev interfaces with conductance
(\ref{eq:Andreev}) in series.

When we close the two normal metal leads of the NSN junction to a ring,
we recover the geometry of the embedding method where the correlated system
is formed by the superconductor. It is therefore interesting to see how
one can recover the linear conductance from this approach.

\subsection{Persistent current and conductance of a superconductor}

For a one-channel ring consisting of a normal conducting region of length
$L_\mathrm{N}$ and a superconducting region of length $L_\mathrm{S}$, the 
solution of the Bogoliubov-de Gennes equation for a boundary condition 
analogous to (\ref{eq:gap}) yields the persistent current 
\cite{Buttiker-Klapwijk,Montambaux}
\begin{equation}
\label{eq:Jsuper}
J(\Phi)=\frac{4}{\pi}\frac{ev_\mathrm{F}}{L_\mathrm{N}+\xi_0 
\tanh(L_\mathrm{S}/\xi_0)}\sum_{m=1}^{\infty}\frac{T_m(X)}{m}\sin(m\Phi)\,.
\end{equation}
Here, $\xi_0=\hbar v_\mathrm{F}/\Delta$ is the superconducting coherence 
length and $T_m(X)$ denotes a Chebyshev polynomial in the variable
\begin{equation}
\label{eq:X}
X=\frac{\cos(k_\mathrm{F}L)}{\cosh(L_\mathrm{S}/\xi_0)}\,.
\end{equation}
In the limit $\xi_0\to\infty$, one obtains a normal conducting ring of length
$L=L_\mathrm{N}+L_\mathrm{S}$ with the persistent current
\begin{equation}
\begin{aligned}
J^0(\Phi) =& \frac{2}{\pi}\frac{ev_\mathrm{F}}{L}\sum_{m=1}^{\infty}\frac{1}{m}
\left[\sin\big(m(\Phi-k_\mathrm{F}L)\big)\right.\\
&\hphantom{\frac{2}{\pi}\frac{ev_\mathrm{F}}{L}\sum_{m=1}^{\infty}\frac{1}{m}
\bigg[}+\left.\sin\big(m(\Phi+k_\mathrm{F}L)\big)\right]\,.
\end{aligned}
\end{equation}
Apart from a factor of two accounting for the spin, this expression reduces
to (\ref{eq:jfluxodd}) or (\ref{eq:jfluxeven}) for $|t|=1$ depending on the
parity of the number of particles per spin. We note, however, that the
expression (\ref{eq:Jsuper}) for the NS ring can, in general, not be expressed
in the form (\ref{eq:jfluxodd}) or (\ref{eq:jfluxeven}).

According to (\ref{eq:transpc}), the dimensionless conductance $g$ can
be obtained from the persistent current at flux $\Phi=\pi/2$
\begin{equation}
\begin{aligned}
J(\pi/2) =& \frac{2}{\pi}\frac{ev_\mathrm{F}}{L_\mathrm{N}+
\xi_0\tanh(L_\mathrm{S}/\xi_0)}\\
&\qquad\times\sum_{m=1}^{\infty}\frac{1}{m}\left\{\sin\!\left[m\left(\frac{\pi}{2}-
\mathrm{Arccos}(X)\right)\right]\right.\\
&\qquad\hphantom{\times\sum_{m=1}^{\infty}\frac{1}{m}\Bigg\{}
+\left.\sin\!\left[m\left(\frac{\pi}{2}+\mathrm{Arccos}(X)\right)\right]\right\}\,.
\end{aligned}
\end{equation}
By means of the Fourier representation of a sawtooth function, one finds
that the absolute value of the persistent current becomes
\begin{equation}
J(\pi/2) = \frac{ev_\mathrm{F}}{L_\mathrm{N}+\xi_0\tanh(L_\mathrm{S}/\xi_0)}\,.
\label{eq:jnsn}
\end{equation}
In view of this result, the superconducting region can be thought of as
a normal-conducting metal of an approximate effective length given by the
minimum of $L_\mathrm{S}$ and $\xi_0$.

It is now straightforward to determine from (\ref{eq:jnsn}) the dimensionless 
conductance
\begin{equation}
\begin{aligned}
g &= \lim_{L_\mathrm{N} \rightarrow \infty}
\left(\frac{J(\pi/2)}{J^0(\pi/2)}\right)^2\\
& = \lim_{L_\mathrm{N} \rightarrow \infty} 
\left(\frac{L_\mathrm{N}+L_\mathrm{S}}{L_\mathrm{N}+\xi_0 
\tanh(L_\mathrm{S}/\xi_0)}\right)^2\,.
\end{aligned}
\end{equation}
Here, the persistent current of the normal ring can again be thought of as 
being obtained from (\ref{eq:jnsn}) in the limit $\xi_0\to\infty$. 
We thus recover the correct result $g=1$ for the dimensionless conductance. 
The leading corrections depend on the ratio 
$[L_\mathrm{S}-\xi_0\tanh(L_\mathrm{S}/\xi_0)]/L_\mathrm{N}$ 
between the relative length of the superconductor, \textit{i.e.} the 
difference between the real length of the superconducting region 
and its effective length, 
and the length of the normal region. Even though here the transmission 
amplitude remains equal to one in the presence of correlations, this example 
gives another demonstratation that the embedding methods works, even for 
having the conductance through a system which is very far to exhibit a Fermi 
liquid behavior.

\end{document}